\documentclass[times,12pt]{article}
\pdfoutput=1
\usepackage{amsmath,amssymb,amsfonts,latexsym,amsthm,enumerate,url}
\usepackage{latex8}
\usepackage{times}
\usepackage{graphicx}
\date{}

\newcommand{\beq}[1]{\begin{equation}\label{#1}}
\newcommand{\enq}[0]{\end{equation}}

\newcommand{\remove}[1]{}

\begin{document}
\title {The Argument against Quantum Computers}

\author{ {\large Gil Kalai}
\\ {\small The Hebrew University of Jerusalem and IDC, Herzliya}\\ }

\maketitle
\begin {center}
{\em Dedicated to the memory of Itamar Pitowsky}
\end {center}

\begin {quotation}
\noindent
My purpose in this paper is to examine, in a general way, the relations between
abstract ideas about computation and the performance of actual physical computers. 
\end {quotation}
\begin {flushright}
Itamar Pitowsky  --  
{\em The physical Church thesis and physical computational complexity}  1990
\end {flushright}

\begin{abstract}
We give a computational complexity argument against the feasibility 
of quantum computers. We identify a very low complexity class of 
probability distributions described by 
noisy intermediate-scale quantum computers, and explain why it will allow neither  
good-quality quantum error-correction nor a demonstration of ``quantum supremacy.''
Some general principles governing the behavior of noisy quantum systems are derived. 
Our work supports the ``physical Church thesis'' studied by Pitowsky (1990) 
and follows his vision
of using abstract ideas about computation to study the performance of actual physical computers. 
\end{abstract}




\section {Introduction}

My purpose in this paper is to 
give an argument against the feasibility of quantum computers.
In a nutshell, our argument is based on the following statement:

\begin {quotation}
Noisy quantum systems will not allow building quantum error-correcting 
codes needed for quantum computation.
\end {quotation}



To support this statement we study the relations between
abstract ideas about computation and the performance of actual physical computers, which are
referred to as noisy intermediate-scale quantum computers or NISQ computers for short.
We need to explain why it is the case that quantum error-correcting codes are needed
to gain the computational advantage of quantum computing,
and we also need to explain why quantum error-correcting codes
are out of reach. The crux of the matter lies in a parameter referred to as the rate of noise.
We will explain why reducing the rate of noise to the level needed for good
quantum error-correction will already enable NISQ computers to demonstrate ``quantum supremacy.''
We will also explain why NISQ computers represent a very low level computational complexity class
which cannot support quantum supremacy. This provides a good, in-principle, 
argument for the infeasibility of both good
quantum error-correcting codes and quantum supremacy.

We will show where the argument breaks down for classical computation. Rudimentary classical
error-correcting codes {\it are supported} by the low-level computational complexity class that describes
NISQ computers. 

Pitowsky (1990)         
formulates and studies a strong form
of the ``physical'' Church--Turing thesis (referred to later as the extended Church--Turing thesis (ECTT))
\footnote{Pitowsky's paper attributes the physical Church--Turing thesis to Wolfram (1985).}, namely,

\begin {quotation}
\noindent
``whether we can invoke physical laws to reduce a computational
problem that is manifestly or presumably of
exponential complexity, and actually complete it in polynomial time.''
\end {quotation}

\noindent
and considers the then recently discovered model
of quantum computers\footnote {Quantum computers were proposed by Feynman (1982), 
Deutsch (1985), 
and others. The idea can be traced to Feynman's 1959 visionary lecture, ``There is Plenty of Room at the
Bottom.''}  as a possible counterexample to this newly formulated physical Church--Turing
thesis. Very strong support for ``quantum supremacy'' -- the ability of quantum computers 
to perform certain tasks that classical computers cannot perform, -- came four 
years after Pitowsky's paper, 
with Peter Shor's discovery (Shor 1994) 
that quantum computers can factor integers efficiently.

Our argument, which is in agreement with quantum mechanics,
supports the validity of the extended Church--Turing thesis but does not rely on this thesis.  
Our argument 
is based on computational complexity
considerations and thus manifests the vision of Pitowsky 
of using
abstract ideas about computation to draw conclusions on the performance of actual physical computers,
and of integrating computational complexity into theories of nature. 
Our computational complexity considerations refer to very low-level computational classes and 
rely on definite mathematical theorems;  
they do not depend on
the unproven conjectures underlying our view of the computational 
complexity world, such as the ${\bf P} \ne {\bf NP}$ conjecture.
The novelty as well as a certain weakness
of my argument is that it uses asymptotic insights (with unspecified constants) to draw conclusions
on the behavior of small- and intermediate-scale systems and, in particular, on the value of  
constants that naively 
appear to reflect just engineering capabilities.

The argument against quantum computers leads to various predictions on near-term experiments,
and to general 
principles governing the behavior of noisy quantum systems.
It may shed light on a variety of questions 
about quantum systems and computation. We emphasize that the argument
predicts the failure of near-term experimental goals of many groups around the world to 
demonstrate on NISQ computers 
quantum supremacy and good-quality quantum error-correcting codes.

In Section 2 
we will consider basic models of computation and
basic insights about computational complexity.
Our argument against quantum computers is presented and discussed in Section 3, 
along with concrete predictions on near-term experimental efforts. The argument crucially relies   
on the study by Kalai and Kindler (2014) 
of noise stability and sensitivity for systems of non-interacting bosons.
This study is built on the theory  of noise stability and noise sensitivity of Boolean functions 
developed by Benjamini, Kalai, and Schramm (1999). 
Section 4 
discusses underlying principles behind  
the failure of quantum computers and some consequences. Section 5 concludes and 
Section 6 is about Itamar.
This paper 
complements, updates, and presents in nontechnical language 
some of my earlier works on the topic of quantum computers; see Kalai (2016, 2016b, 2018),
and earlier papers cited there.

\section {Basic models of computation}

Pitowsky (1990) described three types of
computers where each type is more restrictive than the previous one: 
``the Platonist computer,'' ``the constructivist computer,'' and ``the finitist computer.'' 
We will consider versions of the last two types. Our definitions rely
on the notion of Boolean circuits.\footnote{The precise relation between 
the Turing machine model 
(considered in Pitowsky 1990) 
and the circuit model is interesting and subtle 
but we will not discuss it in this paper.}

\subsection {Pitowsky's ``constructivist computer,'' Boolean functions, and Boolean circuits}

\begin {quotation}
\noindent
The ``constructivist computer''
is just a (deterministic) Turing machine, and associated with
this notion is the class of all Turing machine computable, i.e., recursive functions.
Can a physical machine compute a non-recursive function? 
\end {quotation}
\begin {flushright}
Itamar Pitowsky (1990)  
\end {flushright}

\medskip

{\bf Model 1 -- Computers (circuits):}
A computer (or circuit) has $n$ bits, and it can perform certain logical operations on them.
The NOT gate, acting on a single bit, and the AND gate, acting on two bits, suffice for
the full power of classical computing. 

The outcome of the computation is the value of certain $k$ 
output bits. When $k=1$ the circuit computes a {\it Boolean function},
namely, a function $f(x_1,x_2,\dots,x_n)$ of $n$ Boolean variables ($x_i=1$ or $x_i=0$) so that 
the value of $f$ is also $0$  or $1$.

\subsection {Easy and hard problems and Pitowsky's ``finitist computer''}

\begin {quotation}
\noindent
 The ``finitist computer,'' is a deterministic, polynomial-time Turing
 machine (with a fixed finite number of tapes). Associated with this notion
 is the class of all functions which are computable in a number of steps
 bounded by a polynomial in the size of the input. As is well known, many
 important computational problems seem to lie outside the class {\bf P},
 typically the so-called {\bf NP}-hard problem.
\end {quotation}
\begin {flushright}
Itamar Pitowsky  (1990) 
\end {flushright}

While general circuits can compute arbitrary 
Boolean functions, matters change when we require that the size of the circuit be at most
a polynomial in the number of input variables. Here the size of the circuits is the total number of gates. 

\medskip

{\bf Model 2 -- Polynomial size circuit:} A circuit with $n$ input bits of size at most $A n^c$. 
Here $A$ and $c$ are positive constants.

\medskip

The complexity class {\bf P} refers to 
problems that can be solved  using a polynomial
number of steps in the size of the input.
The complexity class {\bf NP} refers to problems whose solution can be verified in 
a polynomial number of steps.
Our understanding of the computational complexity world depends on a
whole array of conjectures: {\bf NP $\ne$ P} is the most famous one. 
Many natural computational 
tasks are {\bf NP}-complete. (A task in {\bf NP} is {\bf NP}-complete if a computer 
equipped with a subroutine for solving this task can solve, in a polynomial number of steps,
{\it every} problem in {\bf NP}.)  

Let me mention two fundamental examples of computational ``easiness'' and ``hardness.'' 
The first example is: multiplication is easy, factoring is hard!\footnote {Factoring belongs 
to {\bf NP} $\cap$ {\bf coNP} and therefore, 
according to the basic conjectural view of computational 
complexity, it is not an {\bf NP}-complete problem. 
The hardness of factoring is a separate important computational complexity 
conjecture.}   
Multiplying two $n$-digit numbers requires a simple algorithm with a number of steps
that is quadratic in $n$. On the other hand, the best-known algorithm for factoring an $n$-digit 
number to its prime factors is (we think) exponential in $n^{1/3}$.

\begin{figure}
\centering
\includegraphics[scale=0.4]{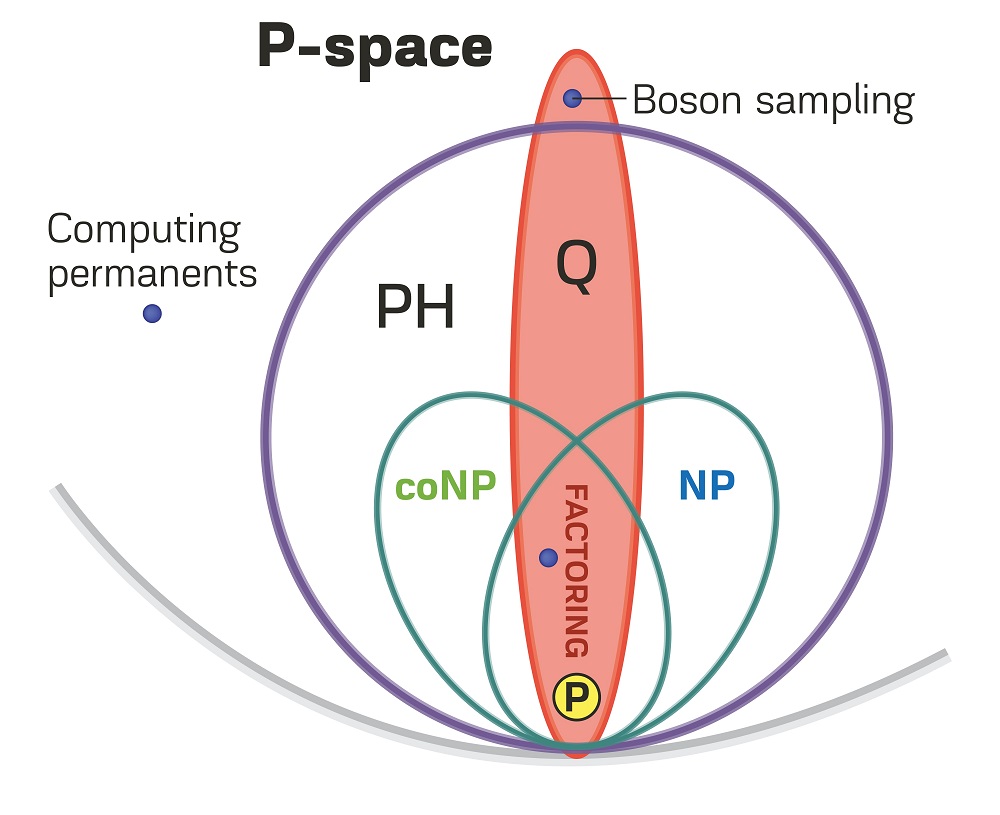}
\caption{The (conjectured) view of some main computational complexity classes. 
The red ellipse represents
efficient quantum algorithms. Note the position of three important 
computational tasks: factoring, computing 
permanents, and boson sampling}
\end{figure}

The second fundamental example of complexity theory is that 
``determinants are easy but permanents are hard.'' 
Recall that the permanent of an $n$-by-$n$ matrix $M$ has a similar (but simpler) formula compared 
to the more famous determinant. 
The difference is that the signs are eliminated, and this difference accounts for 
a huge computational difference. 
Gaussian elimination gives a polynomial-time algorithm 
to compute determinants. By contrast, computing permanents is 
harder than solving {\bf NP}-complete problems.\footnote{The hardness of computing 
permanents defines an important computational class {\bf \#P} 
(in words, number-P or sharp-P) that is larger than {\bf PH} 
(the entire ``polynomial hierarchy'').} 
  



Classical circuits equipped with random bits lead to randomized algorithms,
which are both practically useful and theoretically important.

\medskip

{\bf Model 3 -- Randomized circuit:} A circuit with $n$ bits of input and with one 
additional type of gate that provides a random bit.
Again we assume that the circuit is of size at most $A n^c$ (here, again,  $A$ and $c$
are positive constants). This time the output is a sample from a probability distribution on 
strings of length $k$ of zeroes and ones.\footnote {For simplicity, we use in this paper the term
{\bf P} (rather than {\bf BPP}) 
to describe the computational power 
of polynomial-time computers (and polynomial-size circuits) with randomization. 
We also use the term {\bf Q} 
rather than {\bf BQP} (for decision problems) and {\bf Quantum Sampling} (for sampling problems)   
to describe the computational power of polynomial-time 
quantum computers (and polynomial-size quantum circuits).}

\subsection {Quantum computers}

\noindent
We will now give a brief description (based on circuits) of quantum computers.  

\medskip

{\bf Model 4 -- Quantum computers:}   

\begin {itemize}
\item
A {\it  qubit} is a piece of quantum memory.
A qubit is a unit vector in $\mathbb C ^2$, and   
the state of a qubit is a unit vector in a two-dimensional complex Hilbert space $H = \mathbb C^2$.
The memory of a quantum computer (quantum circuit) consists of $n$ qubits and the 
state of the computer is a unit vector 
in the $2^n$-dimensional Hilbert space, i.e.,  $\mathbb (C^2)^{\otimes n}$.

\item 
{\it A Quantum gate} is a unitary transformation.  We can put one or two qubits through gates 
representing unitary transformations acting on the 
corresponding two- or four-dimensional Hilbert spaces. 
As for classical computers, there is a small list of 
gates that are sufficient for the full power of quantum computing.

\item
{\it Measurement}: Measuring the state of $k$ qubits leads to a probability distribution on $0-1$ 
vectors of length $k$.
\end {itemize}

Quantum computers allow sampling from probability distributions well beyond the capabilities of classical 
computers (with random bits). Shor's famous algorithm shows that quantum computers can 
factor $n$-digit integers efficiently, in roughly $n^2$ steps! It is conjectured  
that quantum computers cannot solve {\bf NP}-complete problems and this 
implies that they cannot compute permanents.

For the notions of noisy quantum computers and for 
fault-tolerant quantum computation that we discuss
next, it is necessary to 
allow measurement throughout the computation process and to be 
able to feed the measurement results back into 
later gates.

\subsection {Noisy quantum circuits}

Next we consider noise. Quantum systems are inherently noisy; we cannot accurately
control them, and we cannot accurately describe them. In fact, every interaction of a quantum
system with the outside world amounts to noise.

\medskip

{\bf Model 5 -- Noisy quantum computers:}
A noisy quantum circuit has the property that every qubit is corrupted in every ``computer cycle''
with a small probability $t$, and every gate is $t$-imperfect.  
Here, $t$ is a small constant called the rate of noise.

\medskip

Here, in a ``computer cycle'' we allow several non-overlapping gates to perform in parallel. 
We do not specify the precise technical meaning of ``corrupt'' and ``$t$-imperfect,'' but the 
following intuitive explanation (for a restricted form of noise called depolarizing noise) 
could be useful. When a qubit is corrupted then its state is replaced by a uniformly distributed 
random state on the same Hilbert state. A gate is $t$-imperfect if 
with probability $t$ the state of the 
qubits that are involved in the gate is replaced by a uniformly 
random state in the associated Hilbert space.

\medskip

{\bf Theorem 1: ``The threshold theorem.''} 
If the error rate is small enough, noisy quantum circuits 
allow the full power of quantum computing.
\medskip

The threshold theorem was proved around 1995 by 
Aharonov and Ben-Or (1997), 
Kitaev (1997), 
and 
Knill, Laflamme, and  Zurek (1998). 
The proof relies on quantum error-correcting codes first introduced 
by Shor (1995) 
and 
Steane (1996). 

A common interpretation of the threshold theorem is that it shows that 
large-scale quantum computers are possible in 
principle. 
A more careful
interpretation is that if we can control noisy intermediate-scale quantum systems well enough, 
then we can build large-scale universal quantum computers. As we will see, there are good 
reasons for why we {\it cannot} control the quality of 
noisy intermediate-scale quantum systems well enough.

\subsection {Quantum supremacy and NISQ devices}

\medskip

\noindent
{\bf Model 6 -- Noisy intermediate-scale quantum (NISQ) computers:}
These are simply noisy quantum circuits with at most 500 qubits.

\medskip

The number 500 sounds arbitrary but there is a reason to regard NISQ circuits 
and other NISQ devices as an important separate computation model: 
NISQ circuits are too small to apply quantum error-correction for 
quantum fault-tolerance, since a single good-quality ``logical'' qubit 
requires a quantum error-correcting code described by a NISQ circuit with  
more than a hundred qubits.

A crucial theoretical and experimental challenge is to understand NISQ 
computers.\footnote{John Preskill coined the terms ``quantum supremacy'' (in 2012) 
and ``NISQ systems'' (in 2018). Implicitly, these notions already play a role  
in Preskill (1998). 
The importance of multi-scale analysis of the quantum computer 
puzzle is emphasized in Kalai (2016).}
Major near-future experimental efforts are aimed at demonstrating quantum 
supremacy using NISQ circuits
and other NISQ 
devices, 
and are also aimed at using NISQ circuits to build 
high-quality quantum error-correcting codes. 

For concreteness we mention three major near-term 
experimental goals. 
\begin {enumerate}

\item [{\bf Goal 1:}]
Demonstrate quantum supremacy via systems of non-interacting bosons. (See Section \ref {s:bs}.)

\item [{\bf Goal 2:}]
Demonstrate quantum supremacy on random circuits (namely, circuits 
that are based on randomly choosing the computation process, in advance), with 50--100 qubits.

\item [{\bf Goal 3:}]
Create distance-5 surface codes on NISQ circuits that require a little over 100 qubits. 

\end {enumerate} 

In our discussion we will refer 
to the task of 
demonstrating with good accuracy boson sampling for 10--20 bosons as {\bf baby goal 1}, 
to the task of 
building with good accuracy 
random circuits with 10--30 qubits, as {\bf baby goal 2}, and 
to the task of 
creating distance-3 surface codes on NISQ circuits that require a little over 20 qubits, 
as {\bf baby goal 3}. 
The argument presented in the next section asserts that attempts to reach 
Goals 1-3 will fail, and the difficulties will be evident already in the baby goals.

\section {The argument against quantum computers}
\label {s:a}

\subsection {The argument}


In brief, the argument against quantum computers is based on a computational 
complexity argument for why NISQ computers cannot demonstrate 
quantum supremacy and cannot create 
the good-quality quantum error-correcting codes needed for quantum computation. 
We turn now to a detailed description of the argument, which 
is based on the following assertions on NISQ devices.

\begin {itemize}

\item [(A)] Probability distributions described (robustly) by NISQ devices can be 
described by low-degree polynomials (LDP). LDP-distributions represent 
a very low-level computational complexity class well inside bounded-depth (classical) computation.

\item [(B)] Asymptotically low-level computational devices cannot lead to superior computation.

\item [(C)] Achieving quantum supremacy is easier than achieving quantum error-correction.

\end {itemize}

Concretely, we argue that Goals 1 and 2 both represent 
superior computation that cannot be reached by the 
low-level computational NISQ devices and that 
Goal 3 (creating distance-5 surface codes) is even 
harder than Goal 2.

\begin{figure}
\centering
\includegraphics[scale=0.5]{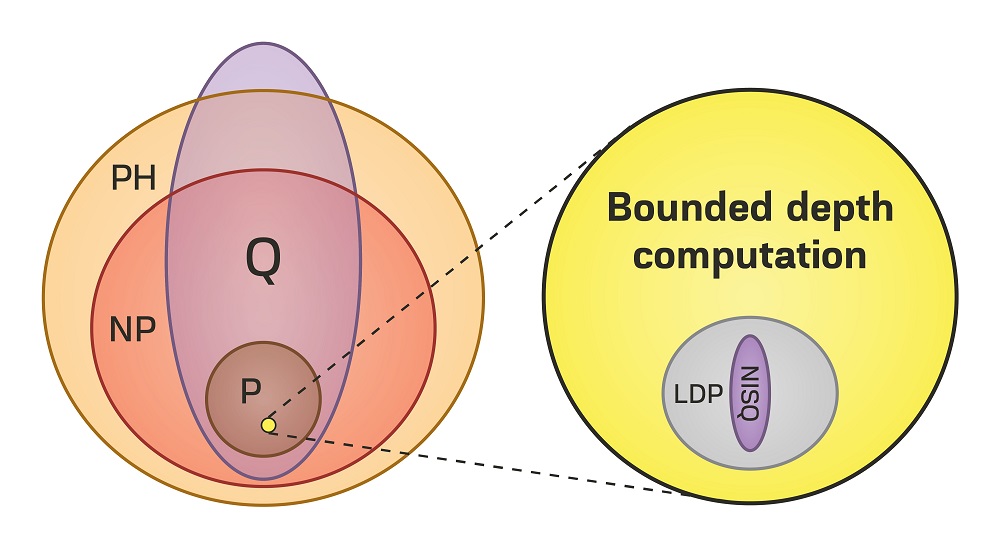}
\caption{NISQ circuits are computationally very weak and therefore 
unlikely to create quantum error-correcting codes needed for quantum computers.}
\end{figure}

We will discuss in Sections \ref {s:bs} and \ref {s:qc} 
our argument for assertion (A) based on the theory of noise sensitivity 
and noise stability, and, in particular, a detailed study of noisy
systems of noninteracting bosons. It would be interesting to explore other 
theoretical arguments for (A).

Assertion (B) requires special attention and it can be 
regarded as both a novel and a weak link of our argument. There 
is no dispute that we can apply asymptotic computational insights to the behavior of computing 
devices  in the small and intermediate scale 
when we know or can estimate the constants involved. This is not 
the case here. The constants depend on (unknown) engineering abilities to control the noise.  
I claim that (even when the constants are unknown) the low-level 
asymptotic behavior implies or strongly 
suggests limitations on the computing power and hence on the engineering ability. 
Moreover,  these limitations already apply for the intermediate scale, namely, for NISQ 
circuits, already when the number of qubits is rather small.
In my view, this type of 
reasoning is behind many important successes regarding 
the interface between the theory of computing and the practice of 
computing.\footnote {A nice example concerns 
the computation of the Ramsey number $R(k,r)$, which is the smallest 
integer $n$ such that in every party 
with $n$ participants you can  find  either a set of $k$ participants 
where every two shook hands, or a set
of $r$ participants where no two shook hands. This is an example where 
asymptotic computational complexity 
insights into the large-scale behavior (namely, when $k$ and $r$ tend to infinity) 
explain the small-scale behavior  
(namely, when $k$ and $r$ are small integers). 
It is easy to verify that $R(3,3)=6$, and it was possible 
with huge computational efforts (and human ingenuity) 
to compute the Ramsey number $R(4,5)=25$ 
(McKay and Radziszowski 1995). 
It is commonly believed that computing $R(7,7)$ and certainly $R(10,10)$ is and will 
remain well beyond our computational ability. 
Building gates that are  two orders of 
magnitude better than the best currently existing gates can be seen as analogous to 
computing $R(10,10)$: we have good computational 
complexity arguments that both these tasks are beyond reach.} 
Several researchers disagree with this part of my analysis and find it unconvincing.

There is substantial empirical evidence for (C) and 
experimentalists often refer to achieving quantum 
supremacy as a step toward achieving quantum error-correction. Quantum 
circuits are expected to exhibit 
probability distributions beyond the power of classical 
computers already for 70--100 qubits, while 
quantum error-correcting codes of the quality needed for quantum 
fault-tolerance will require at least 150--500 qubits. 
It will be interesting to conduct a further theoretical study of the 
hypothesis that the ability to build good-quality quantum error correction codes already 
leads to quantum systems that demonstrate quantum supremacy. 
There is  good theoretical evidence for a weaker statement (C'). 
(Claims (A), (B), and (C') already support a weaker form of the argument 
against quantum computers.)

\begin {itemize}
\item [(C')] High-quality quantum error-correcting codes are not supported by 
the very low-level computational power {\bf LDP} of NISQ systems. 
\end {itemize}

\subsection {Predictions on NISQ computers}
\label {s:pred}

The quality of individual qubits and gates is the major factor for the quality 
of the quantum circuits built from them. One consequence of our 
argument  
is that there is an upper bound on the quality of qubits and gates, which is quite 
close, in fact, to what people can achieve today. 
This consequence seems counterintuitive to many researchers. 
As a matter of fact, the quantum-computing analogue of Moore's law, known as  ``Schoelkopf's law,'' 
asserts, to the contrary, that roughly every three years, quantum decoherence can be delayed by a 
factor of ten.
Our argument implies  that Schoelkopf's law will be broken before we reach the quality 
needed for quantum supremacy and
quantum fault-tolerance. (Namely, now!)\footnote{Schoelkopf's law was formulated by 
Rob Schoelkopf's colleagues based 
on Schoelkopf's experimental breakthroughs, and only later was adopted as a prediction 
for the future. An even bolder related prediction by Neven asserts that quantum computers are gaining 
computational power relative to classical ones at a ``doubly exponential'' rate. Our argument 
implies that Neven's law does not reflect reality.} 
Our first prediction is thus that

\begin {itemize}
\item [(a)] The quality of qubits and gates cannot be improved beyond a certain threshold that is 
quite close to the best currently existing qubits and gates. This applies also to topological qubits 
(Section \ref{s:scope}). 
\end {itemize}



Our argument leads to several further predictions on near-term experiments, for example, on  
distributions of 0-1 strings based on a random quantum circuit (Goal 2), or on
circuits aimed at creating good-quality quantum error-correcting 
codes such as distance-5 and even distance-3 surface codes (Goal 3).

\begin {itemize}

\item [(b)] For a larger amount of noise, robust experimental outcomes are possible but 
they will represent probability distributions that can be expressed in terms of 
low-degree polynomials ({\bf LDP}-distributions, for short) 
that are far away from the desired noiseless distributions.

\item [(c)] For a wide range of smaller amounts of noise, the 
experimental outcomes will not be robust. 
This means that the resulting probability distributions will strongly depend on 
fine properties of the noise and hence the outcomes will be chaotic. 

In other words, in this range of noise rate the probability distributions 
are still far away from the desired noiseless distributions and, moreover,  
running the experiment 
twice will lead to very different probability distributions. 
\end {itemize}

Following predictions (b) and (c) we further expect that 

\begin {itemize}

\item [(d)] 
The effort required to control $k$ qubits 
to allow good approximations of the desired distribution 
will increase exponentially with $k$ and will fail in a fairly small number of qubits. 
(My guess is that the number will be  $\le 20$.)

\end {itemize}

\subsection {Non-interacting bosons}
\label {s:bs}

Caenorhabditis elegans (or C. elegans for short) is a species of a free-living roundworm  
whose biological study shed much light on more complicated organisms. 
Our next model, {\it boson sampling} can be seen as the C. elegans of quantum computing. 
It is both technically and conceptually simpler 
than the circuit model and yet it allows definite and clear-cut insights 
that extend to more general models and more complicated situations. 

\medskip

{\bf Model 7 --  Boson 
Sampling (Troyansky and Tishby 1996; Aaronson and Arkhipov 2013):} 
Given a complex $n$-by-$m$ matrix $X$ with orthonormal rows,
sample subsets of columns (with repetitions) according to the 
absolute value-squared of permanents.
This task is referred to as boson sampling.

\medskip


Boson sampling represents a very limited form of quantum computers based on non-interacting bosons.
(The number of bosons is $n$, and each boson can be in $m$ modes.) 
Quantum circuits can perform boson sampling efficiently on the nose! 
There is a good theoretical argument by Aaronson--Arkhipov (2013) 
that these tasks are beyond the reach of classical computers. 
If we consider non-interacting fermions rather than bosons, we obtain 
a similar sampling task, called 
fermion sampling, with determinants 
instead of permanents. This sampling task can be 
performed by a classical randomized computer (Model 3).  


\medskip

{\bf Model 8 -- Noisy boson sampling (Kalai and Kindler 2014):}  
Let $G$ be a complex Gaussian $n$-by-$m$ noise matrix (normalized so that the expected row norm is 1). 
Given an input matrix $A$, we average the boson sampling distributions over 
$\sqrt{(1-t)}A + \sqrt t G.$ Here, $t$ is the rate of noise.

\medskip

To study noisy boson sampling we expand the outcome distribution in terms of 
Hermite polynomials in $nm$ variables
which correspond to the entries of the input matrix $A$. 
The effect of the noise is exponential decay of high-degree terms and, as it turns out, 
the Hermite expansion for the boson sampling model is very simple and quite beautiful (much like 
the simplicity of the graph of synaptic connectivity of C. elegans).  
This analysis leads to the following theorems:

\medskip

{\bf Theorem 2 (Kalai and Kindler 2014):} When the noise level is constant, 
distributions given by noisy boson sampling are well approximated by their 
low-degree Fourier--Hermite expansion. (Consequently, these distributions can be 
approximated by bounded-depth polynomial-size circuits.)

\medskip

{\bf Theorem  3 (Kalai and Kindler 2014):} 
When the noise level is higher than $1/n$, noisy boson sampling is very sensitive to noise, 
with a vanishing correlation between the noisy distribution and the ideal distribution.

\medskip

Theorem 2 asserts that noisy intermediate-scale systems of non-interacting bosons 
are, for every fixed level of noise,  
 ``classical computing machines''; namely, they can be simulated by classical computers. 
We do not expect {\it classical} computing devices to exhibit quantum supremacy. 
In this case, the argument is especially strong in view 
of the very low-level computational class described by Theorem 2, as well as in 
view of Theorem 3 that asserts that in a wide range of subconstant levels
of noise the experimental outcomes will be unrelated to the noiseless outcomes. 
Furthermore, Theorem 3 strongly suggests that in this wide range of subconstant 
noise levels, experimental outcomes will be chaotic; namely, the correlation between the outcomes 
of two different runs of the experiment will also tend to zero.

\begin{figure}
\centering
\includegraphics[scale=0.5]{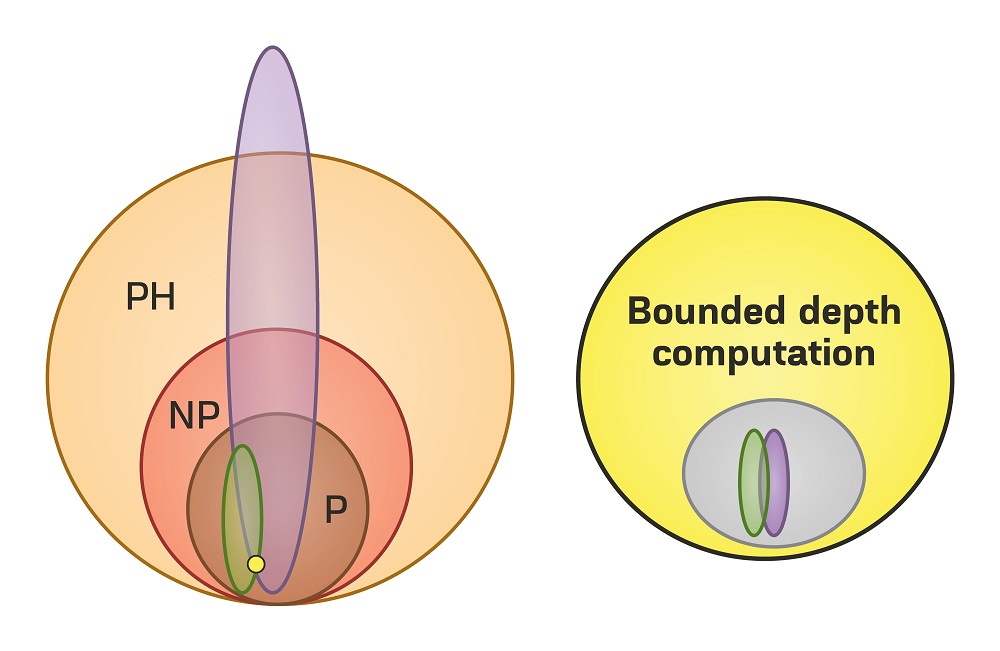}
\caption{The huge computational gap (left) between boson sampling (purple) and fermion sampling (green)
vanishes in the noisy versions (right).}
\label{fig:3}
\end{figure}

\subsection {From boson sampling to NISQ circuits} 
\label {s:qc}

The following open-ended mathematical conjecture is crucial to part (A) of our analysis, 
as well as to predictions (b)--(d) 
in Section \ref {s:pred}.

\medskip

{\bf Conjecture 4:} Theorems 2 and 3 both extend to all NISQ computers
(in particular, to noisy quantum circuits) and to all realistic forms of noise.

\begin {enumerate}
\item [(i)] 
When the noise level is constant, distributions given by NISQ  systems are well approximated
by their low-degree Fourier expansion. In particular, these 
distributions are very low-level computationally.

\item [(ii)] For a wide range of lower noise levels, NISQ systems are very sensitive to noise,
with a vanishing correlation between the noisy distribution and the ideal distribution.
In this range, the noisy distributions depend on fine parameters of the noise.
\end {enumerate}

Assertion (A) of our argument is supported by Conjecture 4(i). 
The conjecture asserts that, for every fixed level of noise, 
probability distributions described by NISQ circuits 
can be simulated by classical computers. As in the case of boson sampling, the argument 
is especially strong in view 
of the very low-level computational class described by Conjecture 4(i), 
and in view of Conjecture 4(ii) that asserts that in a wide range of subconstant levels
of noise the experimental outcomes will be unrelated to the 
noiseless outcomes and, furthermore,  
the correlation between the outcomes of two different runs of the 
experiment will also tend to zero. 

One delicate conceptual aspect of the argument is that it is a multi-scale argument. 
The threshold theorem asserts that noisy quantum circuits for a small enough level of 
noise {\it support} universal computation! However, in the NISQ regime the 
situation is different and Conjecture
4(i) asserts that NISQ circuits are ``classical'' and of a very 
low level of computational complexity
even in the classical computational kingdom. It follows that in the NISQ regime quantum 
supremacy (e.g., Goal 2) is out of reach, and high-quality quantum 
error-correction (e.g. Goal 3) is therefore out of reach as well. 
This behavior of NISQ circuits implies lower bounds on the rate of noise
that show that, in larger scales, the threshold theorem is not relevant.

Results by Gao and Duan (2018), and by 
Bremmer, Montanaro, and Shepherd (2017)  
give some support to Conjecture 4(i).

\subsection {The scope of the argument}

\label {s:scope}

NISQ computers account for most of the effort to create stable qubits via 
quantum error-correction, 
to demonstrate quantum supremacy, and to 
ultimately build universal quantum computers. 
We expect (Section \ref {s:Pns}) that noise stability and the 
very low-level computational class LDP that we identified for noisy NISQ systems
apply in greater generality.

Specifically, it is plausible that the failure of NISQ computers to achieve stable logical qubits 
extends to systems, outside the NISQ regime, that do not apply quantum error-correction. 
It is plausible that Google's attempts to build ``surface codes'' 
with quantum circuits of 50--1000 qubits 
will meet the same fate as Microsoft's attempts to build a single surface-code qubit 
based on non-Abelian anyons. 
The microscopic quantum process for Microsoft's attempts is not strictly in the NISQ regime, 
but it is plausible that 
it represents the same primitive computational power of noisy NISQ 
circuits.

In brief, the extended argument against quantum computers is based on a paradox.
On the one hand, you cannot achieve quantum supremacy without quantum error-correction. 
On the other hand,
if you could build quantum error-correcting codes of the required 
quality then you could demonstrate quantum supremacy
directly on smaller and simpler quantum systems that involve no 
quantum error-correction. To sum

\begin {quotation}
Noisy quantum systems that refrain from using error correction represent a low-level computational
ability that will not allow building quantum error-correcting 
codes needed for quantum computation.
\end {quotation}

It is not easy to define quantum devices that ``refrain from using error correction,'' but 
this property holds for NISQ circuits that simply do not have enough qubits to 
use quantum error correction, and appears to hold for all known experimental processes
aimed at building high quality physical qubits including toplogical qubits. What we see next is 
that the low level complexity class of NISQ circuits does allow building basic forms of 
classical error-correcting codes.    

\subsection {Noise stability, noise sensitivity, and classical computation}
\label {s:ns}

The study of noisy boson sampling relies on a  study of 
Benjamini, Kalai, and Schramm (1999)  
of noise stability 
and noise sensitivity of Boolean functions.
See 
Kalai (2018) where both these theories and their connections 
to statistical physics, voting rules,  
and computation are discussed. A crucial mathematical aspect 
of the theory is that the noise has a simple 
description in terms of Fourier expansion: it reduces  ``high frequency'' 
coefficients exponentially and, therefore, 
noise stability reflects ``low frequency'' (low-degree) Fourier expansion. 

The theory of noise stability and noise  sensitivity 
posits a clear distinction between classical information and quantum information.
The class {\bf LDP} of probability distributions that can be approximated by 
low-degree polynomials does not support quantum supremacy and quantum error-correction 
but it still supports robust {\it classical} information, and 
with it also classical communication and computation.
The (``majority'') Boolean function, defined by
$f(x_1,x_2,\dots,x_n) = 1$ if more than half the variables have a value of 1, 
allows for very robust bits
based on a large number of noisy bits or qubits and
admits excellent low-degree approximations. 
It can be argued that {\it every  form} of robust information, communication,  
and computation in nature  
is based on a rudimentary form of classical error-correction where 
information is encoded by repetition (or simple variants of repetition) 
and  decoded by the majority function (or a simple variant of it).  
In addition to this rudimentary form of classical error-correction,
we sometimes witness more sophisticated forms of classical error-correction.

\subsection{The extended Church--Turing thesis revisited} 

The extended Church--Turing thesis asserts (informally) that

\medskip

(ECTT) It is impossible to build computing devices that demonstrate computational supremacy.

\medskip

Here, computational supremacy is the ability to perform certain computations 
far beyond the power of digital computers. 

A {\it classical computing device} (or a {\bf P}-device) is a computing device for which 
we can show or have good reason to believe that it can be modeled by 
polynomial-size Boolean circuits. 

We now formulate the weak extended Church--Turing thesis that reads

\medskip

(WECTT) It is impossible to build a {\it classical} computing device 
that demonstrates computational supremacy.

\medskip

WECTT seems almost tautological and it is in wide agreement.
A popular example of an application of WECCT is to the human brain. 
As many argue, it is unlikely  that 
quantum effects in the human brain lead to some sort of quantum computational supremacy.  
Another implication of WECCT is that the recent proposal by Johansson and Larsson (2017) 
to demonstrate Shor's factoring algorithm on certain classical 
analog devices cannot be realistic.  


The argument given in this section asserts that for constant level of noise, NISQ devices 
are {\bf P}-devices. Consequently, WECTT suffices to rule out quantum supremacy of NISQ devices. 
What makes this argument stronger is that we have good evidence that

\begin {enumerate}

\item
Probability distributions described by NISQ devices for constant error rate 
are actually modeled by {\bf LDP}, a computational class well below 
bounded-depth computation, which is well below {\bf P}.
\item
The outcomes of the computation process are chaotic for 
a wide range of subconstant levels of errors. 

\end {enumerate}

We argue that NISQ devices are, in fact,  
low-level classical computing devices, and I regard it as strong evidence that engineers will not be able 
to reach the level of noise required for quantum supremacy. Others argue that 
the existence of low-level simulation (in the NISQ regime) for every fixed level of noise 
does not have any bearing on the engineering question of reducing the level of noise.
These sharply different views will be tested in the next few years.

In his comments on this paper, physicist and friend Aram Harrow 
wrote that my predictions are so sweeping that they say that hundreds of different research teams 
will all stop making progress. 
Indeed, my argument and predictions say that the goals 
of these groups (such as Goals 1--3 and other, more ambitious, goals) will not be achieved.  
Rather, if I am correct, these research teams will make 
important progress in our understanding of the inherent limitations of computation in nature, 
and the limitations of human ability to control quantum systems.

Aram Harrow also wrote: 
``The Extended Church--Turing Thesis (ECTT) is a major conjecture, 
implicating the work of far more than quantum computing. 
Long before Shor's algorithm it was observed that quantum systems are hard in 
practice to simulate because 
we do not know algorithms that beat the exponential scaling and work in all cases, 
or even all practical cases.  
Simulating the 3 or so quarks (and gluons) of a proton 
is at the edge of what current supercomputers can achieve despite 
decades of study, and going to larger nuclei (say atomic weight 6 or more) is still out of reach. 
If the extended Church--Turing thesis is actually true then there is a secret poly-time simulation 
scheme here that the lattice QCD community has overlooked.  Ditto for chemistry where 
hundreds of millions of dollars are spent each year simulating molecules.  
Thus the ECTT is much stronger than just saying we cannot build quantum computers.''

I agree with the spirit of Harrow's comment, if not with the specifics,  
and the comment can serve as a nice transition to the 
next section that discusses a few underlying principles and consequences that follow 
from the failure of quantum computing.

\section {The failure of quantum computers: Underlying principles and consequences}

In this section we will propose some general underlying principles 
for noisy quantum processes that follow from the failure of quantum computers. 
(For more, see Kalai 2016b and Kalai 2018.)


\subsection {Noise stability of low-entropy states} 

\subsection *{Principle 1: Probability distributions described
by low-entropy states are noise stable and can be expressed by low-degree polynomials.}

\label {s:Pns}

The situation for probability distributions described by 
NISQ systems appears to be as follows: noise causes the 
terms, in a certain Fourier-like expansion, 
to be reduced exponentially with the degree. Noise-stable states, namely, those 
states for which the effect of the noise is small, 
have probability distributions expressed by low-degree terms. 
Probability distributions realistically (and robustly) obtained by NISQ devices 
represent a very low-level computational complexity 
class,  {\bf LDP}, the class of probability distributions 
that can be approximated by low-degree polynomials. 
In the absence of quantum fault-tolerance this description 
extends to (low-entropy) quantum states that 
arbitrary quantum circuits and (other quantum devices) can reach.

Noise stability and descriptions by low-degree polynomials might  
be relevant to
some general aspects of quantum systems in nature. 
In the previous section we considered one such aspect: the feasibility of 
classical information and computation. 
The reason that our argument against quantum computation does not apply for classical 
computation is that noise stability still allows basic forms of 
classical error-correction. In fact, every  form of robust information, 
communication, and computation 
in nature is based on some rudimentary form of classical error-correction 
that is supported by low-degree polynomials. 
We will consider two additional aspects: the possibility of efficiently learning  
quantum systems, 
and the ability of physical systems to reach ground states. 

Before we proceed I would like to make the following two remarks. 
The first remark is that the specific ``Fourier-like expansion'' 
required for the notions of noise stability and noise sensitivity is different for different situations. 
(The computational class {\bf LDP} applies to all these situations.) 
An interesting direction for future 
research would be to extend the noise stability/noise sensitivity 
dichotomy and to find relevant Fourier-like expansions and definitions of noise  
for mathematical objects of high-energy physics. The second remark is that 
the study of noise stability is important in other scientific disciplines, and the 
assumption that systems are noise stable (aka ``robust'') 
may provide useful information for studying them. Robustness is an important ingredient in the study of 
chemical and biological networks; see, e.g., Barkai and 
Leibler (1997).


\subsubsection*{Learnability}
It is a simple and important insight of machine learning that the 
approximate value of a low-degree polynomial
can efficiently be learned from examples.
This offers an explanation for our ability to understand
natural processes and the parameters defining them.
Rudimentary physical processes with a higher entropy rate may represent simple extensions
of the class of {\bf LDP} for which efficient learnability is still possible, 
and this is an interesting topic for further research.

\subsubsection* {Reaching ground states}
Reaching ground states is computationally hard ({\bf NP}-hard) for classical
systems, and even harder for quantum systems.  
So how does nature reach ground states so often? The common answer relies on two ingredients:
the first is that physical systems operate in 
positive temperature rather than zero temperature, and
the second is that nature often reaches meta-stable states rather than ground states.
However, these explanations are incomplete: we have good theoretical reasons to think that,
for general processes in positive temperature,
reaching meta-stable states is computationally intractable as well.
First, for general quantum or classical systems,
reaching a meta-stable state can be
just as computationally hard as reaching a ground state.
Second, one of the biggest breakthroughs in computational complexity,
the ``PCP theorem,'' 
asserts (in physics language) that positive temperature offers no 
computational complexity relief for general
(classical) systems. 

Quantum evolutions and states 
approximated by low-degree polynomials represent 
severe computational restrictions, and the results on the hardness of 
reaching ground states 
no longer apply. This may give a theoretical support to the 
natural phenomenon of easily reached ground states.


\begin{figure}
\centering
\includegraphics[scale=0.6]{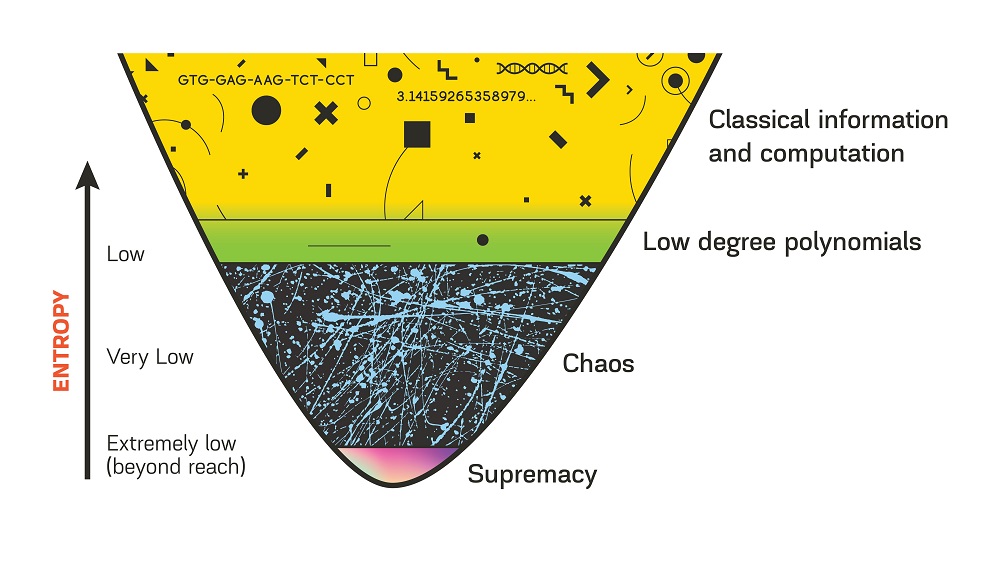}
\caption{Low-entropy quantum states give probability distributions described  by low degree polynomials, 
and very low-entropy quantum states give chaotic behavior. Higher entropy enables classical information.}
\label{fig:4}
\end{figure}

\subsection {Noise and time-dependent evolutions}

\subsection *{ Principle 2: Time-dependent (local) quantum evolutions are inherently noisy.}

\medskip

It is an interesting challenge to give a satisfactory mathematical 
description of what ``time dependent'' means. We proposed in Kalai (2016) 
to base a 
lower bound on the rate of noise  
on a measure of the non-commutativity
of unitary operations. This measure of non-commutativity can also serve as an ``internal'' clock
of the noisy quantum evolution. We also proposed in Kalai (2016) a 
restricted class of noisy evolutions called ``smoothed Lindblad evolutions'' 
defined via a certain smoothing operation 
that mathematically expresses the idea that in the absence of quantum error-correction, 
noise accumulates. 
An important consequence of Principle 2 is that 
there are inherent lower bounds on the quality of 
qubits, or, more concretely,  a universal upper bound on the number of non-commuting Pauli 
operators you can apply to a qubit before it get destroyed. 
We note that lower bounds on the rate of noise in terms of a 
measure of non-commutativity 
arise also in Polterovich (2014) in the study of quantizations in symplectic geometry.



\subsection {Noise and correlation}

\subsection* 
{Principle 3: Entangled qubits are subject to positively
correlated noise.}

\medskip

Entanglement, the quantum analog of correlations, is a 
notion of great importance in quantum physics and 
in quantum computing, 
and the famous {\it cat state} 
 ${\frac{1}{\sqrt 2}}\left|00\right\rangle +{\frac{1}{\sqrt 2}} \left|11\right\rangle$
represents the simplest example of entanglement and the strongest form of entanglement between two qubits.
Principle 3 asserts that the errors for the two qubits of a cat state necessarily 
have large positive correlation. This is an observed 
and accepted phenomenon for gated qubits and without
quantum error-correction it is inherited to all pairs of entangled qubits.
{\it Error synchronization} refers to a substantial probability 
that a large number of qubits, much beyond the average rate of noise, are corrupted.
An important consequence of Principle 3 is that 
complicated states and evolutions lead to error-synchronization.
Principle 3 leads to additional predictions about NISQ systems that can be 
tested for current experiments.

\begin {itemize}
\item [(e)] Every pair of gated qubits 
will be subject to errors with large positive correlation. 

\item [(f)] Every pair of qubits in a cat state 
will be subject to errors with large positive correlation. 

\item [(g)] All experimental attempts to achieve Goals 2 and 3 will 
lead to a strong effect of error-synchronization.
\end {itemize}

We emphasize that our argument against quantum computers in Section \ref {s:a}
does not depend on Principle 3 and, moreover, the main trouble with quantum computers 
is that the noise rate of qubits and gates cannot be reduced toward the threshold for 
fault-tolerance. 
Correlation is part of this picture since, as is well known, 
the value of the threshold becomes smaller when there is a 
larger positive correlation for 2-qubit gated errors. 
We note (Kalai 2016) that Prediction (f) implies that the quality of logical qubits 
via quantum error-correction 
is limited by the quality of gates.  
See also Chubb and Flammia (2018) 
for an interesting analysis of the effect of correlated errors on surface codes.

A reader may question how it is possible 
that entangled states will experience correlated noise, even 
long after they have been created and even if the constituent particles 
are far apart and not directly interacting.    
We  emphasize that Principle 3 does not reflect any form of 
``non local'' noise and that no ``additional'' correlated 
noise on far apart entangled qubits is ever assumed or conjectured. 
Principle 3  simply reflects two facts. The first fact is that errors 
for gated qubits are positively correlated from the time of creation. 
The second fact is that unless the error rate is small enough to allow 
quantum error-correction 
(and this may be precluded by the argument of the previous section), 
cat states created indirectly  
via a sequence of computational steps will inherit errors with large positive correlation.

{\bf Remark:} 
Error-synchronization is related to an interesting possibility (Kalai 2016b) 
that may challenge 
conventional wisdom and 
can also be checked for 
 near term NISQ (and other) experiments. 
A central property of Hamiltonian models 
of noise in the study of quantum fault-tolerance  
(see, e.g., Preskill 2013),   
is that error fluctuations are sub-Gaussian. Namely,
when there are $N$ qubits, the standard deviation
for the number of qubit errors behaves like $\sqrt N$
and the probability of more than $t \sqrt N$ errors 
decays as it does for Gaussian distribution.
Many other models in mathematical physics have a similar property. 
However, it is possible that fluctuations for the total amount of errors 
for noisy quantum circuits and other quantum systems with interactions
are actually super-Gaussian and perhaps even linear. 

\subsection {A taste of other consequences}

The failure of quantum computers and the infeasibility of high-quality 
quantum error-correction have 
several others far-reaching consequences imposing severe 
restrictions of humans' and nature's ability 
to control quantum states and evolutions. 
Following the tradition of using cats for quantum 
thought experiments, consider an ordinary living cat. 

If quantum computation is not possible, 
it will be impossible
to teleport the cat;
it will be impossible
to reverse time in the life of the cat;
it will be impossible
to implement the cat
on a very different geometry;
it will be impossible
to superpose the lives  of two distinct cats;
and, finally, even if we place the cat in an isolated
and monitored environment,
the life of this cat cannot 
be predicted.  

These 
restrictions will already be in place  
when our ``cat'' represents realistic (but involved) quantum evolutions described 
by quantum circuits in the small scale, namely,  
with less than 50 qubits (baby goals 1--3 can serve as good candidates for such ``cats'').
See Kalai (2016b) for a detailed discussion. 

\begin{figure}
\centering
\includegraphics[scale=0.3]{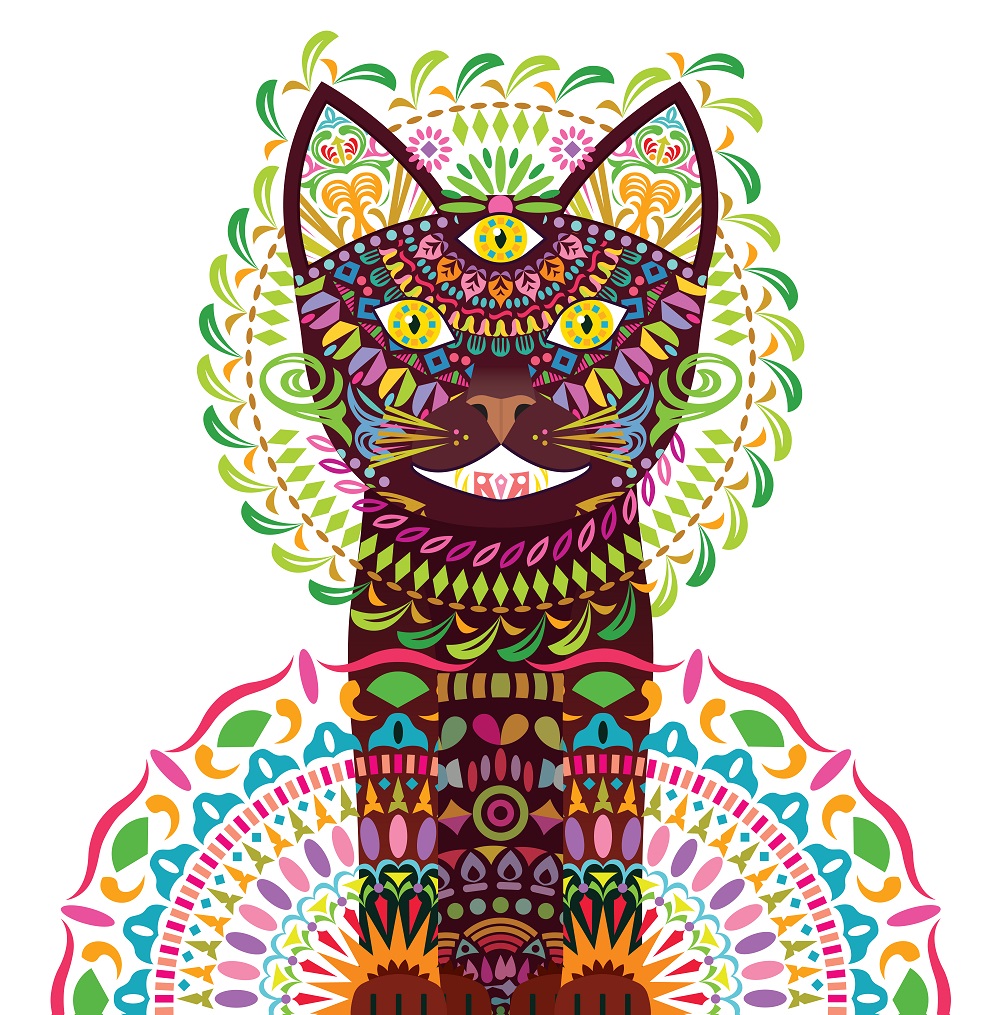}
\caption{ Human imagination allows for a cat with an unusual 
geometry. Drawing by Netta Kasher (a tribute to Louis Wain)}
\label{fig:cat}
\end{figure}

\section {Conclusion}

\begin {quotation}
\noindent
The intrinsic study of computation transcends human-made artifacts,
and its expanding connections and interactions with all sciences,
integrating computational modeling, algorithms, and complexity into theories
of nature and society, marks a new scientific revolution!
\end {quotation}
\begin {flushright}
Avi Wigderson -- {\em Mathematics and Computation} 2019.
\end {flushright}


Understanding noisy quantum systems and potentially even the failure of 
quantum computers is related to the fascinating 
mathematical theory of noise stability and noise sensitivity 
and its connections to the theory of computing. Our study shows how  
insights from the computational complexity
of very low-level computational classes  support  the extended Church--Turing thesis.
Exploring this avenue may have important implications 
for various areas of quantum physics. These connections between 
mathematics, computation, and physics are characterized 
by a rich array of conceptual and technical methods and points of view. 
In this study, we witness a delightfully thin gap  
between fundamental and philosophical issues on the one hand 
and  practical and engineering aspects on the other.

\section {Itamar}
\label {s:ita}

This paper is devoted to the memory of Itamar Pitowsky.
Itamar was great. He was great in science and great in the humanities.
He could think and work like a mathematician, and like a physicist,
and like a philosopher, and like a philosopher of science,
and probably in various additional ways. And he enjoyed
the academic business greatly, and took it seriously, with humor.
Itamar had an immense human
wisdom and a modest, level-headed way of expressing it.


Itamar's scientific path  interlaced with mine 
in many ways. Itamar  made important contributions to the foundations of 
the theory of computation, probability theory, and quantum mechanics, 
all related to some of my own 
scientific endeavors and to this paper. Itamar's
approach to the foundation of quantum mechanics
was that
quantum mechanics is a theory of non-commutative
probability that (like classical probability)
can be seen as a mathematical language for the other laws of
physics. (My work, to a large extent, adopts this point of view.
But, frankly, I do not quite understand the other points of view.)

\begin{figure}
\centering
\includegraphics[scale=0.65]{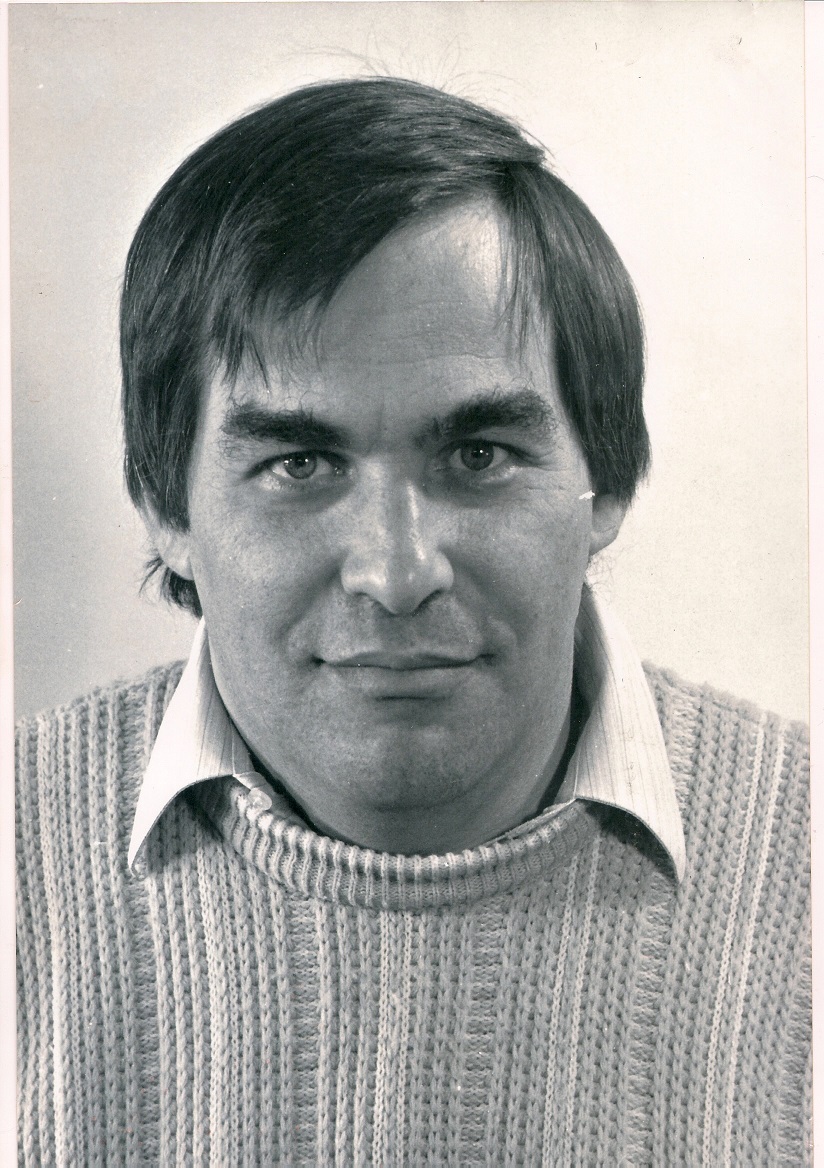}
\caption{Itamar Pitowsky around 1985}
\label{fig:6}
\end{figure}

In the late `70s when Itamar and I were both graduate
students I remember him 
enthusiastically sharing his thoughts on the Erd\H{o}s--Tur\'an problem with me.
This is a mathematical conjecture that asserts
that if we have a sequence of integers
$0<a_1<a_2< \cdots < a_n< \cdots$ such that the
infinite series $\sum \frac {1}{a_n}$ diverges then we can
find among the elements of the sequence
an arithmetic progression of length $k$, for every $k$. 
Over the years both of us spent some time on this conjecture
without much to show for it. This famous conjecture is still open even for $k=3$.
Some years later
both Itamar and I became interested, for entirely different reasons,
in remarkable geometric objects called cut polytopes.
Cut polytopes are obtained by
taking the convex hull of characteristic vectors of
edges in all cuts of a graph. Cut polytopes
arise naturally when you try to understand
correlations in probability theory and Itamar wrote several fundamental papers
studying them; see  Pitowsky (1991). 
Cut polytopes came into play, in an
unexpected way, in the work
of Jeff Kahn and myself 
where we disproved Borsuk's conjecture (Kahn and Kalai 1993). 
Over the years, Itamar and I  became interested in
Arrow's impossibility theorem (Arrow 1950),  
which Itamar regarded
as a major 20th-century intellectual achievement. He gave a course centered
around this theorem and years later so did I. The study of the noise stability and 
noise sensitivity of Boolean functions has close ties to Arrow's theorem.

The role of skepticism in science and how (and if)
skepticism should be practiced
is a fascinating issue. It is, of course,
related to this paper that describes skepticism
over quantum supremacy and quantum fault-tolerance, widely believed to
be possible. Itamar and I  had long
discussions about skepticism in science, and about the nature and practice of scientific debates,
both in general and in various specific cases. This common interest was well suited to
the Center for the Study of Rationality at the Hebrew University that we were both members of, 
where we and other friends
enjoyed many discussions, and, at times, heated debates.

A  week before Itamar passed away, Itamar,  Oron Shagrir, and I sat at our little CS
cafeteria and talked about probability. Where does probability come from?
What does probability mean? Does it just represent human uncertainty? Is it just an emerging mathematical concept that
is convenient for modeling? Do matters
change when we move from classical to quantum mechanics? When we move to quantum physics the
notion of probability itself changes for sure, but is there a change in the interpretation of
what probability is?  A few people passed by and listened, and it felt like this was a direct continuation of
conversations we had had while
we (Itamar and I; Oron is much younger) were students in the early `70s.
This was to be our last meeting.

\subsection*{Acknowledgement}
Work supported by ERC advanced grants 320924, \& 834735, 
NSF grant DMS-1300120, 
and BSF grant 2006066. 
The author thanks Yuri Gurevich, 
Aram Harrow, 
and Greg Kuperberg for helpful comments, and Netta Kasher for drawing Figures 1--5.


\begin{thebibliography}{99}

{\small




\bibitem 
{AarArk13} S. Aaronson and A. Arkhipov,
The computational complexity of linear optics, {\it Theory of Computing} 4 (2013), 143--252.


\bibitem {AhaBen97} D. Aharonov and M. Ben-Or,
Fault-tolerant quantum computation with constant error, STOC '97,
ACM, New York, 1999, pp. 176--188.


\bibitem {Arr50} K. Arrow, A difficulty
in the theory of social welfare, {\it Journal of Political
Economy} 58 (1950) , 328--346.


\bibitem {BarLei97} 
N. Barkai, and S. Leibler. Robustness in simple 
biochemical networks, {\it Nature,} 387 (1997), 913.

\bibitem{BKS99}
I.~Benjamini, G.~Kalai, and O.~Schramm,
\newblock Noise sensitivity of Boolean functions and applications to
percolation,
\newblock {\em Publications Math\'ematiques de l'Institut des Hautes \'Etudes 
Scientifiques} 90 (1999), 5--43.

\bibitem {BMS17} M. J. Bremner, A. Montanaro, and D. J. Shepherd, Achieving quantum supremacy with 
sparse and noisy commuting quantum computations, {\em Quantum} 1, 8 (2017). 


\bibitem {ChuFla18}  C. T. Chubb and S. T. Flammia, 
Statistical mechanical models for quantum codes with correlated noise, arXiv:1809.10704.

\bibitem {Deu85}
D. Deutsch, Quantum theory, the Church--Turing principle and the
universal quantum computer, {\it Proceedings of the Royal Society of London A} 
400 (1985), 96--117.

\bibitem {Fey82} R. P. Feynman, Simulating physics with computers,
{\it International Journal of Theoretical Physics} 21 (1982), 467--488.

\bibitem {GaoDua18} X. Gao and L. Duan, 
Efficient classical simulation of noisy 
quantum computation, arXiv:1810.03176.

\bibitem {JohLar17} N. Johansson and J.-A. Larsson, Realization of Shor's algorithm at 
room temperature, arXive:1706.03215.

\bibitem {KahKal93} J. Kahn and G. Kalai, A counterexample to Borsuk's conjecture,
{\it Bulletin of the American Mathematical Society} 29 (1993), 60--62.

\bibitem {Kal16} G. Kalai, The quantum computer puzzle, {\it Notices 
of the American Mathematical Society} 
63 (2016), 508--516.

\bibitem {Kal16b} G. Kalai, The quantum computer puzzle (expanded version), arXiv:1605.00992.

\bibitem {Kal18} G. Kalai, Three puzzles on mathematics, computation and 
games, {\it Proceedings of the International 
Congress of Mathematicians} 2018, Rio de Janeiro, Vol. I, pp. 551--606.

\bibitem{KalKin14} G. Kalai and G. Kindler, Gaussian noise sensitivity 
and BosonSampling, arXiv:1409.3093.


\bibitem{Kit97} A.~Y.~Kitaev, Quantum error
correction with imperfect gates, in {\it Quantum Communication,
Computing, and Measurement 
}, Plenum Press, New York, 1997, pp. 181--188.


\bibitem{KLZ98} 
E.~Knill, R.~Laflamme, and W.~H.~Zurek, Resilient
quantum computation: Error models and thresholds, {\it Proceedings 
of the Royal Society of London A }{454} (1998), 365--384.

\bibitem {McKRad95} B. D. McKay and S. P. Radziszowski, R(4,5)=25, {\it Journal of 
Graph Theory} 19 (1995), 309--322.


\bibitem{Pit90}
I. Pitowsky, The physical Church thesis
and physical computational complexity, {\it lyuun, A Jerusalem
Philosophical Quarterly} 39 (1990), 81--99.


\bibitem {Pit91}
I. Pitowsky, Correlation polytopes: Their geometry and complexity,
{\it Mathematical Programming} A50 (1991), 395--414.

\bibitem {Pol14} L. Polterovich,  Symplectic geometry of quantum
noise, {\it Communications in Mathematical Physics} 327 (2014), 481--519.

\bibitem {Pre98} J. Preskill, Quantum computing: Pro and con,
{\it Proceedings of the Royal Society of London A} 454 (1998), 469--486.


\bibitem{Pre13} J. Preskill,
Sufficient condition on noise correlations for scalable quantum computing, 
{\it Quantum Information and Computing} 13 (2013), 181--194.


\bibitem {Sho94} P. W. Shor, Polynomial-time
algorithms for prime factorization and
discrete logarithms on a quantum computer,
{\it SIAM Rev.} 41 (1999), 303--332.
(Earlier version,
{\it Proceedings of the 35th Annual Symposium on Foundations of
Computer Science}, 1994.)

\bibitem {Sho95} P. W. Shor, Scheme for reducing
decoherence in quantum computer
memory, {\it Physical Review A} 52 (1995), 2493--2496.


\bibitem {Ste96}
A.~M.~Steane, Error-correcting codes in
quantum theory, {\it Physical Review Letters} 77 (1996), 793--797.


\bibitem {TroTis96} L. Troyansky and N. Tishby, Permanent uncertainty: On the quantum evaluation
of the determinant and the permanent of a matrix, in {\it Proceedings of the 4th Workshop on Physics and Computation}, 1996.


\bibitem {Wig19} A. Wigderson, {\em Mathematics and Computation,} Princeton University Press, 2019.


\bibitem {Wol85} S. Wolfram, Undecidability and intractability
in theoretical physics, {\it Physical Review Letters} 54 (1985), 735--738.



}


\end{thebibliography}
\end {document}